\documentclass[prd,twocolumn,superscriptaddress,showpacs,preprintnumbers,amsmath,amssymb]{revtex4}
\usepackage[dvipdfmx]{graphicx}
\usepackage{dcolumn}
\usepackage{bm}
\usepackage[dvipdfmx]{color}

\usepackage{color}

\begin{document}
\title{Perfect Abelian dominance of quark confinement in SU(3) QCD}
\author{Naoyuki~Sakumichi}
\affiliation{Theoretical Research Division, Nishina Center, RIKEN, Wako, Saitama 351-0198, Japan}
\author{Hideo~Suganuma}
\affiliation{Department of Physics, Kyoto University, Kitashirakawaoiwake, Sakyo, Kyoto 606-8502, Japan}
\date{\today}
\begin{abstract}
We study the Abelian projection of quark confinement in SU(3) quenched lattice QCD, in terms of the dual superconductor picture.
In the maximal Abelian gauge, we perform the Cartan decomposition of the non-Abelian gauge field on a $32^4$ lattice with spacing $a \simeq0.058, 0.10$ fm  (i.e., $\beta =6.4, 6.0$), and investigate the interquark potential $V(r)$, the Abelian part $V_{\rm Abel}(r)$, and the off-diagonal part $V_{\rm off}(r)$. 
For the potential analysis, we use both on-axis data and several types of off-axis data, with larger numbers of gauge configurations. 
Remarkably, we find almost perfect Abelian dominance of the string tension (quark-confining force) on the large-volume lattice.
Also, we find a simple but nontrivial relation of $V(r) \simeq  V_{\rm Abel}(r) + V_{\rm off}(r)$.
\end{abstract}

\maketitle

\section{Introduction}
\label{sec1}

Deriving quark confinement directly from quantum chromodynamics (QCD) is one of the most important unsolved issues remaining in particle physics \cite{QCD}.
Although its analytical proof is not yet known from QCD, lattice QCD Monte Carlo simulations show that the static quark-antiquark (Q$\bar{\rm Q}$) potential is well reproduced by a sum of the Coulomb and linear confinement terms as \cite{QCD,BSS93,TSNM02}
\begin{equation}
V(r) = - \frac{A}{r}+\sigma r +C,
\label{eq:Cornell}
\end{equation}
with the interquark distance $r$, the string tension $\sigma$, the color-Coulomb coefficient $A$, and an irrelevant constant $C$.
Since quark confinement is phenomenologically interpreted with a linear interquark potential at long distances, the strength of quark confinement is controlled by the string tension $\sigma$, i.e., the linear slope of the interquark potential.
The linear confinement term $\sigma r$ is considered to be caused by ``one-dimensional squeezing'' of the interquark color-electric flux, which is shown by lattice QCD studies \cite{QCD}.
Historically, such a one-dimensional property of hadrons leads to several interesting theoretical frameworks, such as the string theory, the flux-tube picture \cite{CNN79}, and the Lund model \cite{Lund83} for hadron reactions based on the Schwinger mechanism.
Nevertheless, the physical origin of the color-flux squeezing is not yet clearly understood.

\begin{figure}[b]
\centering
\includegraphics[width=8.4cm,clip]{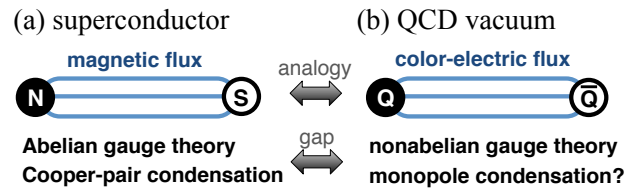}
\caption{
(a) In superconductors, magnetic flux is repelled due to Cooper-pair condensation, 
and is squeezed into a one-dimensional tube like the Abrikosov vortex. 
(b) In the dual-superconductor picture, the QCD vacuum is regarded as an electromagnetic dual version of the superconductor: the interquark color-electric flux is squeezed into a one-dimensional form due to magnetic-monopole condensation.}
\label{Fig:1}
\end{figure}

For the one-dimensional color-flux squeezing, the dual-superconductor picture proposed by Nambu, 't~Hooft and Mandelstam in the 1970s \cite{NtHM} seems to provide a plausible scenario.
In this picture, the QCD vacuum is assumed to be a ``dual superconductor'' (i.e., the electromagnetic dual version of a superconductor), 
and the electric-flux squeezing is caused by the dual version of the Meissner effect in superconductors, similar to the formation of the Abrikosov vortex (see Fig.~\ref{Fig:1}).
This picture provides us with a guiding principle for the modeling of quark confinement.
For example, based on the dual superconductivity, the dual Ginzburg-Landau theory \cite{DGL} is formulated as a low-energy effective model of QCD, and describes confinement phenomena and the flux-tube structure of hadrons.

However, there are two large gaps between the dual superconductor and the QCD vacuum:
\begin{itemize}
\item The dual superconductor is governed by an Abelian U($1$) gauge theory like QED, while QCD is a non-Abelian SU($3$) gauge theory.
\item The dual superconductor requires the condensation of 
color-magnetic monopoles (i.e., the electromagnetic dual version of Cooper pairs), 
while QCD does not have such monopoles as elementary degrees of freedom.
\end{itemize}
On account of these two gaps, it seems difficult to define the dual-superconducting picture precisely from QCD. 
Indeed, the gluon field in QCD has both diagonal and off-diagonal parts, $A_\mu= \sum_{a=1}^8 A_\mu^aT_a =\vec A_\mu\cdot \vec H+ \sum_\alpha A_\mu^\alpha T_\alpha$, with the diagonal generators (Cartan subalgebra) $\vec H = (T_3, T_8)$ and the off-diagonal generators $\{ T_\alpha \}_{\alpha = 1,2,4,5,6,7}$ of SU($3$).
Here, $T_3= {\rm diag}(1/2,-1/2,0)$ and $T_8=(1/2\sqrt{3}) \times {\rm diag}(1,1,-2)$.
The off-diagonal part $\sum_\alpha A_\mu^\alpha T_\alpha$ induces the non-Abelian nature.

As a possible solution to fill in the above two gaps, 't~Hooft proposed the ``Abelian projection'' \cite{tH81,EI82}, a mathematical procedure to reduce QCD to an Abelian gauge theory including monopole degrees of freedom.
In particular, lattice QCD studies \cite{KSW87,SY90,SNW94,AS99} indicate that the maximally Abelian (MA) projection---a special Abelian projection---seems to be successful in extracting infrared-relevant Abelian degrees of freedom from QCD. 
(The concept of Abelian projection may have wide utility, and it was recently applied to multiband superconductors in condensed matter physics \cite{YH13}.)

The MA projection has two steps.
(i) The diagonal part of gluons $\vec A_\mu\cdot \vec H$ is maximized by minimizing the off-diagonal part $\int d^4x\sum_{\mu,\alpha} |A_\mu^\alpha(x)|^2$ in Euclidean SU(3) QCD under the gauge transformation \cite{KSW87,SY90,SNW94,AS99}.
This procedure, called MA gauge fixing, is a partial gauge fixing which remains Abelian gauge degrees of freedom of U(1)$^2$.
(ii) SU(3) QCD is projected onto a U(1)$^2$ Abelian gauge theory by dropping the off-diagonal part of gluons, $\sum_\alpha A_\mu^\alpha T_\alpha$.
It is known that the MA-projected Abelian theory well reproduces QCD phenomena at long distances, which is called ``Abelian dominance" \cite{EI82,SY90,SNW94,AS99,Kondo}. 
As a remarkable fact in the MA gauge, color-magnetic monopoles appear as the topological object corresponding to the nontrivial homotopy group $\pi_2($SU$(3)/$U$(1)^{2})=\mathbb{Z}^2$ \cite{tH81,EI82,Kondo}. 
Thus, by the MA projection, QCD is reduced to an Abelian gauge theory including both electric- and magnetic-monopole currents, which is expected to provide a theoretical basis for the monopole-condensation scheme for the confinement mechanism. 
Several lattice QCD studies show the appearance of monopole worldlines covering the whole system \cite{KSW87,SNW94} and the magnetic screening \cite{SATI00}, which suggest ``monopole condensation''.
Thus, the QCD system resembles a dual superconductor by way of the MA projection.

However, such lattice studies were performed mainly in simplified SU(2) color QCD, and there are only several pioneering studies on the Abelian dominance of quark confinement in actual SU(3) color QCD \cite{STW02,DIK04,L04}.
Stack~{\it et~al.} first studied the Abelian projection of the Q$\bar{\rm Q}$ potential using SU(3) quenched lattice QCD with $10^3 16$ and $16^4$ at $\beta=5.9$ and $6.0$ \cite{STW02}, respectively.
Boryakov~{\it et~al.} investigated similar subjects in SU(3) quenched QCD with $16^3 32$ at $\beta=6.0$ and full QCD, using the simulated annealing algorithm \cite{DIK04}. 
Also in SU(3) color, approximate Abelian dominance of the string tension was found as 
$\sigma_{\rm Abel}/\sigma \simeq 0.83$--$0.93$.
(See Table~\ref{tab1}.)

In this paper, we perform a quantitative analysis for the MA projection of the Q$\bar{\rm Q}$ potential in SU(3) QCD at the quenched level. 
Then, we examine the Abelian dominance of quark confinement. 
After the MA gauge fixing, the SU(3) link variables are factorized with respect to the Cartan decomposition of SU($3$) into U(1)$^2$ and SU(3)$/$U(1)$^2$.
Then, we calculate the original SU(3) Q$\bar{\rm Q}$ potential $V(r)$, the Abelian part $V_{\rm Abel}(r)$, and the off-diagonal part $V_{\rm off}(r)$.
For each sector, we investigate both the linear confinement and Coulomb parts through an accurate fit analysis.

\begin{table}[t]
\caption{The simulation conditions ($\beta$, the lattice size $L^3L_t$, and the gauge configuration number $N_{\rm con}$) 
and the results (lattice spacing $a$, the physical spatial size $La$, and the string tension ratio $\sigma_{\rm Abel}/\sigma$).
Here, the lattice spacing is determined so as to reproduce the string tension of $\sigma = 0.89$ GeV/fm.
The results of previous studies are shown on the last three lines.
We investigate the six types of interquark directions, while Stack {\it et~al.} \cite{STW02} did one type (on-axis) and 
Bornyakov {\it et~al.} \cite{DIK04} did three types [the interquark directions are $(1,0,0), (1,1,0), (1,1,1)$].
}
\label{tab1}
\begin{ruledtabular}
\begin{tabular}{ccclllcc}
$\beta$ & $L^3 L_t$ & $N_{\rm con}$ & \, $a$ [fm] &  $La$ [fm] & \, $\sigma_{\rm Abel}/\sigma$ & $N_{\rm dir}$ & Ref.  \\
\hline
   $6.4$   & $32^4$     & 200    & 0.0582(2)  & 1.86(1) & \, 1.015(09)  &  6 \\
    $6.0$   & $32^4$     & 200    & 0.1022(5)  & 3.27(1) & \, 1.009(10)  &  6 \\
   $5.8$   & $16^3 32$  & 600   & 0.148(1)  & 2.37(2) & \, 1.00(2)  & 6 \\
\hline
   $6.0$   & $16^3 32$  & 600    & 0.102(1)  & 1.64(1) & \, 0.94(1)  &  6 \\
   $6.0$   & $12^3 32$  & 400    & 0.104(1)  & 1.25(4) & \, 0.94(3)  &  6 \\
    $6.2$   & $16^3 32$  & 400   & 0.075(1)  & 1.20(1) & \, 0.95(2)  &  6  \\
\hline
\hline
     $5.9$   & $10^316$  & 440  &0.123(3)  & 1.23(3) & \, 0.93(6)  & 1 & \cite{STW02} \\
     $6.0$   & $16^4$  & 600  &0.105(1)  & 1.68(2) & \, 0.90(4)  & 1 & \cite{STW02} \\
     $6.0$   & $16^3 32$  & -- &0.1 & 1.6 & \, 0.83(3)  & 3 & \cite{DIK04} \\
\end{tabular} 
\end{ruledtabular}
 \end{table}

\begin{figure*}[t]
\centering
\includegraphics[width=17.8cm,clip]{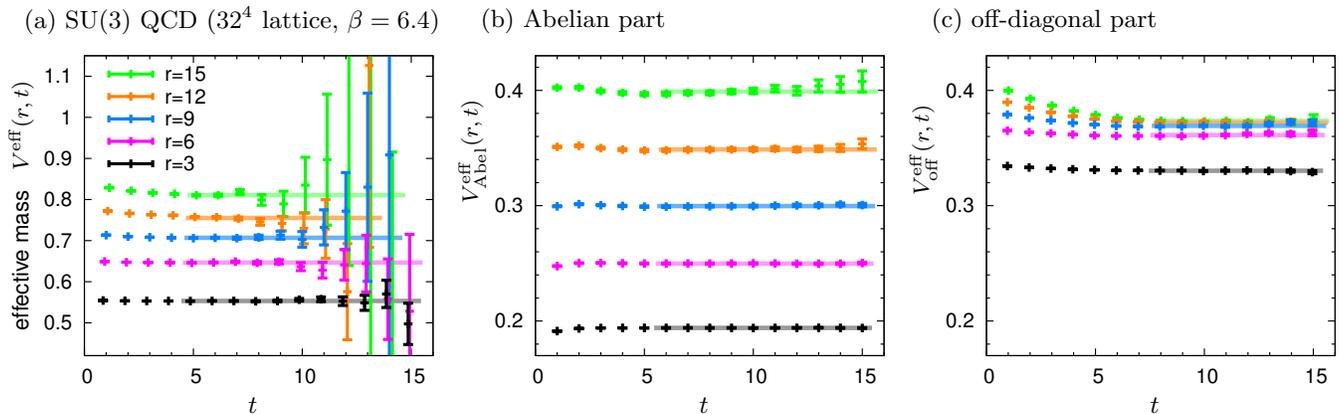}
\caption{Examples of effective mass plots at $\beta =6.4$ on the $32^4$ lattice for (a) SU(3) QCD,  (b) the Abelian part, and (c) the off-diagonal part.
Here, we display on-axis data of $r=3, 6, 9, 12, 15$ in lattice units.
The solid horizontal lines denote the obtained values of $V(r)$, $V_{\rm Abel}(r)$, and $V_{\rm off}(r)$,
from the least-squares fit with the single-exponential form (\ref{sup-eq:exp-fit}),
and are extended in the corresponding fit range of $t_\textrm{min} \leq t \leq t_\textrm{max}-1$.
}
 \label{SupFig:Meff64}
\end{figure*}

\section{Cartan decomposition in SU(3) Maximal Abelian gauge}
\label{sec:2}
In the lattice QCD formalism, the gauge field is described by the link variable $U_\mu(s) =e^{iagA_\mu(s)} \in$ SU(3), with the lattice spacing $a$ and the gauge coupling $g$.
To perform the SU(3) MA gauge fixing, we maximize
\begin{equation}
R_{\rm MA}[U_\mu(s)]\equiv \sum_{s} \sum_{\mu=1}^4  
{\rm tr}\left( U_\mu^\dagger(s)\vec H U_\mu(s)\vec H\right), 
\label{MAgf}
\end{equation}
under the SU(3) gauge transformation 
\begin{equation}
U_\mu(s)\rightarrow U_\mu^\Omega(s)\equiv \Omega(s)U_\mu(s)\Omega^\dagger(s+\hat\mu), 
\end{equation}
with $\Omega(s) \in$ SU(3).
Let $U_\mu^{\rm MA} (s) \in$ SU(3) be the link variables in the MA gauge.
We extract the Abelian part of the link variables 
\begin{equation}
u_\mu(s) = 
\exp \left( i\theta_\mu^3(s) T_3 + i\theta_\mu^8(s) T_8 \right)
\in {\rm U(1)}_3 \times {\rm U(1)}_8,
\end{equation}
by maximizing the norm 
${\rm Re} \, {\rm tr}\left( U_\mu^{\rm MA}(s) u_\mu^\dagger(s) \right)$.
The off-diagonal part of the link variables is defined as
\begin{equation}
M_{\mu}(s) \equiv U_\mu^{\rm MA}(s) \, u_{\mu}^{\dagger }(s)
 \in {\rm SU}(3) / ({\rm U}(1)_3 \times {\rm U}(1)_8),
\end{equation}
which leads to the Cartan decomposition of the SU(3) group, 
\begin{equation}
U^{\rm MA}_\mu(s)=M_\mu(s)u_\mu(s)
=e^{i \sum_\alpha \theta_\mu^\alpha(s) T_\alpha}e^{i\vec \theta_\mu(s) \cdot \vec H}.
\end{equation}

In the MA gauge, there remains the residual ${\rm U(1)}^2$ gauge symmetry 
with the global Weyl symmetry, i.e., the permutation symmetry 
on the color index \cite{STW02,IS99}.
In fact, $R_{\rm MA}$ in Eq.~(\ref{MAgf}) is invariant 
under the ${\rm U(1)}^2$ gauge transformation
\begin{equation}
U_\mu(s)\rightarrow U_\mu^\omega(s)\equiv 
\omega(s)U_\mu(s)\omega^\dagger(s+\hat\mu),
\label{eq:U(1)2Gauge}
\end{equation}
with $\omega(s) \in$ U($1$)$_3 \times$U($1$)$_8$,
and it is also invariant under the global color permutation. 
By the residual gauge transformation (\ref{eq:U(1)2Gauge}), 
$u_\mu(s)$ and $M_\mu(s)$ transform as
\begin{equation}
\begin{split}
u_\mu(s)&\rightarrow u_\mu^\omega(s)\equiv 
\omega(s)u_\mu(s)\omega^\dagger(s+\hat\mu)  \\ 
M_\mu(s)&\rightarrow M_\mu^\omega(s)\equiv 
\omega(s)M_\mu(s)\omega^\dagger(s),
\label{eq:U(1)2Gauge2}
\end{split}
\end{equation}
where $M_\mu(s)$ keeps the form of 
$e^{i \sum_\alpha \theta_\mu^\alpha(s)T_\alpha}\in$ SU(3)/U(1)$^2$.
Then, the Abelian link variables $u_\mu(s)$ behave as gauge variables 
in U(1)$^2$ lattice gauge theory, 
which is similar to the compact QED.
As mentioned above, the MA-projected Abelian theory has not only the electric current, but also the magnetic-monopole current.

\section{Setting for numerics and Cartan decomposition of the Q$\bar{\textbf Q}$ potential}
\label{sec:3}
%
For the numerical analysis, we perform SU(3) quenched lattice QCD Monte Carlo simulations using the standard plaquette action. 
We investigate several $12^3 32$, $16^3 32$, and $32^4$ lattices at $\beta \equiv 6/g^2=5.8$--$6.4$.
For the $12^3 32$ and $16^3 32$ lattices, we identify $12^3$ and $16^3$ as the spatial sizes and $32$ as the temporal one.
The simulation condition and related quantities are summarized in Table~\ref{tab1}.
%
After a thermalization of $20 000$ sweeps, we sample the gauge configuration every $500$ sweeps.
Then, to study the Abelian and off-diagonal parts, we numerically maximize $R_{\rm MA}$ in Eq.~(\ref{MAgf}) for each configuration until it converges.
We use the overrelaxation method for the maximization algorithm to improve convergence \cite{STW02}. 
As for the stopping criterion, we stop the maximization algorithm, 
when the ratio of the deviation $\Delta R_{\rm MA}/R_{\rm MA}$ 
after the one-sweep gauge transformation is less than $\sim 10^{-9}$ for the $12^332$ and $16^332$ lattices and $\sim 10^{-5}$ for the $32^4$ lattices in this calculation.
The converged values of $R_{\rm MA}/(4L^3L_t)$ are
$0.7072(6)$ with $16^332$ at $\beta=5.8$;
$0.7321(11)$, $0.7322(7)$, and $0.7318(3)$ with $12^332$, $16^332$, and $32^4$ at $\beta=6.0$, respectively;
$0.7510(7)$ with $16^332$ at $\beta=6.2$; and
$0.7656(3)$ with $32^4$ at $\beta=6.4$.
%
%
Here, the values in parentheses denote the standard deviation.
Because of the fairly small standard deviation of $R_{\rm MA}$, 
the maximized value of $R_{\rm MA}$ is almost the same over $200$--$600$ gauge configurations. 
In fact, our procedure seems to escape bad local minima, 
where $R_{\rm MA}$ is relatively small. 
Then, we expect that the Gribov copy effect is not 
so significant in our calculation.

\begin{figure*}[t]
\centering
\includegraphics[width=17.8cm,clip]{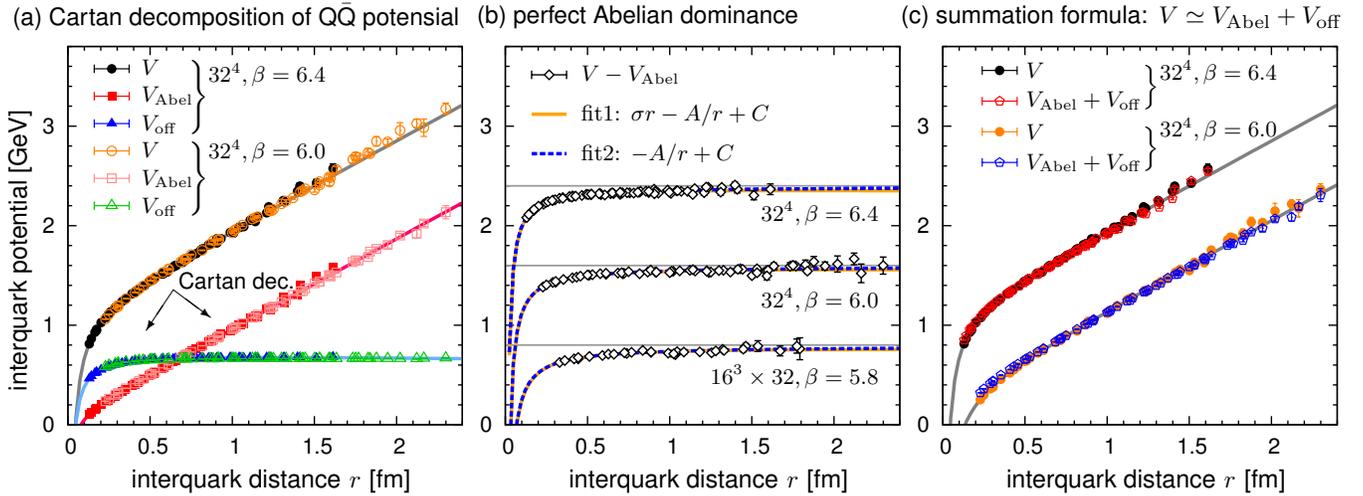}
\caption{(a) Cartan decomposition of the ${\rm Q}\bar{\rm Q}$ potential. 
The circles, squares, and triangles denote the Q$\bar{\rm Q}$ potential $V(r)$, the Abelian part $V_{\rm Abel}(r)$, and the off-diagonal part $V_{\rm off}(r)$, respectively.
The filled and open symbols denote the data at $\beta=6.4$ and $6.0$, respectively.
The curves are obtained by the best fit with Eq.~(\ref{eq:Cornell}) for each part at $\beta=6.4$ as listed in Table~\ref{tab:fit64}.
%
(b) Fit analysis of $V(r) - V_{\rm Abel}(r)$ to illustrate the perfect Abelian dominance of quark confinement.
The orange solid curve is the best fit with the Coulomb-plus-linear ansatz of Eq.~(\ref{eq:Cornell}).
The blue dotted curve is the best fit with the pure Coulomb ansatz [Eq.~(\ref{eq:Cornell}) with $\sigma=0$].
(c) Comparison between $V_{\rm Abel}(r) + V_{\rm off}(r)$ (red open pentagons) and $V(r)$ (black filled circles) at $\beta =6.0$ and $6.4$, except for an irrelevant constant.
Their agreement indicates the summation formula~(\ref{eq:Nontrivial}).}
\label{fig:pot_Dec}
\end{figure*}

\begin{table*}[t]
\caption{Fit analysis with the Coulomb-plus-linear ansatz for the Q$\bar{\rm Q}$ potentials. 
For each potential, the best-fit parameter set $(\sigma, A, C)$ is listed in the functional form of Eq.~(\ref{eq:Cornell}) in lattice units.
For $V - V_{\rm Abel}$, the best-fit parameter set $(A, C)$ is also listed with $\sigma=0$.
}
\label{tab:fit64}
\begin{ruledtabular}
\begin{tabular}{cccc ccc ccc}
 & \multicolumn{3}{c}{$32^4, \beta=6.4$} 
 & \multicolumn{3}{c}{$32^4, \beta=6.0$} 
 & \multicolumn{3}{c}{$16^3 32, \beta=5.8$}  \\
\cline{2-4}\cline{5-7}\cline{8-10}
 &  $\sigma $  & $A$ & $C$ 
 &  $\sigma $  & $A$ & $C$ 
 &  $\sigma $  & $A$ & $C$  \\
\hline
$V$ \ 
    & \,\, 0.01528(12) & 0.265(3) & 0.598(1) 
    & \,\, 0.0471(4) & 0.290(7) & 0.659(4)  
    & \,\, 0.0988(19)      & \, 0.315(25) & \, 0.679(15)  \\
$V_{\rm Abel}$  
    & \,\, 0.01550(06) & 0.056(1)  & 0.167(1)
    & \,\, 0.0475(2) & 0.044(3)  & 0.178(2)
    & \,\, 0.0988(08)      & \, 0.039(10) & \, 0.183(06)  \\
$V_{\rm off}$
    & $-$0.00037(03) & 0.175(1) & 0.391(1)
    & $-$0.0009(1) & 0.139(3) & 0.415(1) \\
\hline
$V - V_{\rm Abel}$
    & $-$0.00024(11)   & 0.209(3) & 0.432(1) 
    & $-$0.0005(3)   & 0.247(6) & 0.481(3) 
    & $-$0.0010(17)   & \, 0.285(21) & \, 0.502(12) \\
$V - V_{\rm Abel}$
    &  0 & 0.205(1) & 0.429(1) 
    &  0 & 0.240(3) & 0.476(1) 
    &  0  & \, 0.273(09) & \, 0.494(03) \\
\hline
$V_{\rm Abel} + V_{\rm off}$
    & \,\, 0.01528(07) & 0.227(2) & 0.556(1) 
    & \,\, 0.0459(2) & 0.201(4) & 0.602(2) \\
\end{tabular}
\end{ruledtabular}
\end{table*}

We calculate the Q$\bar{\rm Q}$ potential
\begin{equation}
V (r) = - \lim_{t\rightarrow \infty}\frac{1}{t}\ln \left\langle 
W_\mathcal{C} \left[ U_\mu(s)  \right] \right\rangle 
\end{equation}
from the Wilson loop
$W_\mathcal{C} \left[ U_\mu(s)  \right]
\equiv  {\rm tr}  \left( \prod_\mathcal{C}  U_\mu (s)  \right) $
for the on-axis and off-axis interquark directions as $(1,0,0), (1,1,0), (1,1,1), (2,1,0), (2,1,1), (2,2,1)$.
Here, $\mathcal{C}$ denotes the closed $r \times t$ rectangle trajectory, 
and $\langle \cdots \rangle$ is the statistical average over the gauge configurations. 
We extract $V(r)$ from the least-squares fit with the single-exponential form
\begin{equation}
\left< W(r,t) \right> = C(r) e^{-V(r) t}.
\label{sup-eq:exp-fit}
\end{equation}
Here, we choose the fit range of $t_\textrm{min} \leq t \leq t_\textrm{max}$ such that the stability of the so-called effective mass
\begin{equation}
V^\textrm{eff}(r,t) \equiv \ln \frac{\left< W(r,t) \right>}{\left< W(r,t+1) \right>}
\label{sup-eq:eff-mass}
\end{equation}
is observed in the range $t_\textrm{min} \leq t \leq t_\textrm{max}-1$.

We also calculate the MA projection of the Q$\bar{\rm Q}$ potential
\begin{equation}
V_{\rm Abel} (r) = - \lim_{t\rightarrow \infty} \frac{1}{t}
\ln \left\langle 
W_\mathcal{C} \left[ u_\mu(s)  \right] \right\rangle 
\end{equation}
from the Abelian Wilson loop in the MA gauge $W_\mathcal{C} \left[ u_\mu(s)\right]$, which is invariant under the residual Abelian gauge transformation~(\ref{eq:U(1)2Gauge2}).
%
%
For the off-diagonal part of the Q$\bar{\rm Q}$ potential
\begin{equation}
 V_{\rm off} (r) = - \lim_{t\rightarrow \infty} \frac{1}{t}
\ln \left\langle 
W_\mathcal{C} \left[ M_\mu(s)  \right] \right\rangle, 
\end{equation}
we fix the residual Abelian gauge symmetry~(\ref{eq:U(1)2Gauge2}) by taking the ${\rm U(1)}_3 \times {\rm U(1)}_8$ Landau gauge, 
which is performed by maximizing
\begin{equation}
 R_L [u_\mu(s)]\equiv \sum_{s} \sum_{\mu=1}^4 
{\rm Re} \, {\rm tr} \left( u_\mu(s) \right)
\end{equation}
under the gauge transformation~(\ref{eq:U(1)2Gauge2}).

To calculate the Q$\bar{\rm Q}$ potentials accurately, 
we adopt the gauge-invariant smearing method, 
which reduces the excited-state components 
and enhances the ground-state overlap in the Q$\bar{\rm Q}$ system.
We adopt the smearing parameter $\alpha =2.3$, which is 
a standard value for the measurement of the Q$\bar{\rm Q}$ potential 
\cite{BSS93,TSNM02}, and 
the iteration number $N_\textrm{smr}=60, 60, 40,20$ for SU(3) QCD 
and the off-diagonal part at $\beta = 6.4, 6.2, 6.0, 5.8$, respectively.
For the Abelian part, we adopt $\alpha =2.3$ and 
the iteration number $N_\textrm{smr}=7, 3, 2, 1$  
at $\beta = 6.4, 6.2, 6.0, 5.8$, respectively.
Here, $N_{\rm smr}$ is chosen so that the ground-state overlap is largely enhanced for each part and at each $\beta$.
%
We have confirmed that the result is almost unchanged 
when the iteration number $N_{\rm smr}$ is changed to some extent.

We exclude the four small-$r$ data points ($r \leq 2$ in lattice units) because they suffer from a large discretization.
(Many previous studies used the same manner, e.g., Refs.~\cite{BSS93,STW02,DIK04}.)
In addition, we exclude some large-$r$ data points because of a lack of statistics for convergence.
We use the jackknife method for the error estimate.
By way of illustration, we show the effective mass plots for each part in lattice units at $\beta=6.4$ for on-axis data in Fig.~\ref{SupFig:Meff64}.
Owing to the smearing, the effective mass seems to be stable even for small $T$.


\section{perfect Abelian dominance of quark confinement and summation formula}
%

Figure~\ref{fig:pot_Dec}(a) shows the lattice QCD results of the original Q$\bar{\rm Q}$ potential $V(r)$, the Abelian part $V_{\rm Abel}(r)$, and the off-diagonal part $V_{\rm off}(r)$ at $\beta =6.4$ and $6.0$.
For each part, we show the best-fit curve of Eq.~(\ref{eq:Cornell}) at $\beta =6.4$.
The fit analyses are summarized in Table~\ref{tab:fit64}.

Remarkably, we find almost perfect Abelian dominance of the string tension, $\sigma_{\rm Abel} \simeq \sigma$, on the $32^4$ lattice at $\beta =6.4, 6.0$ and on the $16^3 32$ lattice at $\beta=5.8$, as summarized in Table~\ref{tab1}.
On the other hand, the off-diagonal part $V_{\rm off}(r)$ has almost zero string tension, $\sigma_{\rm off} \simeq 0$, and is almost a pure Coulomb-type potential.
To demonstrate the perfect Abelian dominance conclusively, we investigate $V(r) - V_{\rm Abel}(r)$ through a fit analysis with the Coulomb-plus-linear ansatz of Eq.~(\ref{eq:Cornell}) (fit 1) and the pure Coulomb ansatz of Eq.~(\ref{eq:Cornell}) with $\sigma =0$ (fit 2).
In Fig.~\ref{fig:pot_Dec}(b) and Table~\ref{tab:fit64}, fit 1 reveals that
$V(r) - V_{\rm Abel}(r)$ has almost zero string tension, $\sigma \simeq 0$.
Moreover, $V(r) - V_{\rm Abel}(r)$ is well described by the pure Coulomb ansatz (fit 2) because fits 1 and 2 are consistent.
Thus, we conclude that there is no difference between the string tensions in $V(r)$ and $V_{\rm Abel}(r)$ with almost perfect precision.

On the smaller lattices of $12^3 32$ and $16^3 32$ at $\beta=6.0$ 
and $16^3 32$ at $\beta=6.2$, however, we cannot see the perfect Abelian dominance of the string tension as shown in Table~\ref{tab1}.
By using the on-axis data points on the $16^4$ lattice at $\beta =6.0$, only approximate Abelian dominance, $\sigma_{\rm Abel}/\sigma \simeq 0.90$, was reported in the previous study by Stack {\it et~al.} \cite{STW02}.
By using the five types of off-axis data with on-axis data, the ratio $\sigma_{\rm Abel}/ \sigma \simeq 0.94$ is improved on $16^3 32$ at $\beta=6.0$ (as shown in Table~\ref{tab1}), but it is not perfect.
Also, we find only approximate Abelian dominance on 
the $12^3 32$ lattice at $\beta =6.0$ and the $16^3 32$ lattice at $\beta =6.2$, as $\sigma_{\rm Abel}/\sigma \simeq 0.94$--$0.95$.
Therefore, we conclude that the perfect Abelian dominance of the string tension can be seen on the $32^4$ lattice at $\beta=6.4, 6.0$ and the $16^3 32$ lattice at $\beta=5.8$, but it cannot seen on the $12^3 32$ and $16^3 32$ lattices at $\beta=6.2, 6.0$.

\begin{figure}[t]
\centering
\includegraphics[width=7.7cm,clip]{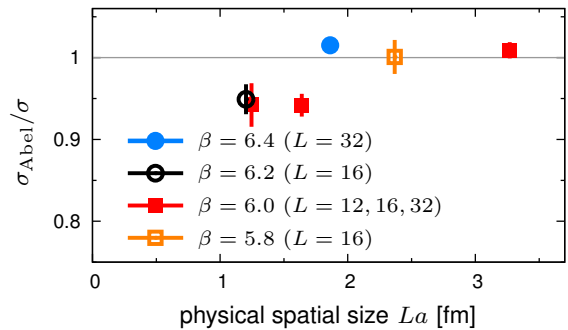}
\caption{Physical spatial-size dependence of $\sigma_{\rm Abel}/\sigma$.
Perfect Abelian dominance ($\sigma_{\rm Abel}/\sigma \simeq 1$) seems to be realized when the spatial size $La$ is sufficiently large.
}
\label{fig:PAD}
\end{figure}

It is likely that a large physical volume is necessary for the perfect Abelian dominance of the string tension.
In fact, as shown in Table~\ref{tab1} and Fig.~\ref{fig:PAD}, the $32^4$ lattice at $\beta=6.4, 6.0$ and the $16^3 32$ lattice at $\beta=5.8$ have large spatial volumes of $La = 32a \simeq 1.9, 3.3$ fm and $La=16a \simeq 2.4$ fm on a side,  while the lattices of $12^3 32$ and $16^3 32$ at $\beta=6.0$ 
and $16^3 32$ at $\beta=6.2$ have slightly 
small spatial volumes of $La \simeq 1.2$--$1.6$ fm.
Thus, perfect Abelian dominance seems to be realized 
when the physical spatial volume of the lattice is approximately 
larger than $(2$ fm$)^3$.
In contrast to Abelian dominance for the long-distance confinement properties, there is a significant difference between $V(r)$ and $V_{\rm Abel}(r)$ at short distances.
Fine lattices are necessary for the accurate analysis of the short-distance behavior of the potential, and it is preferable to use our results obtained on the fine lattice at $\beta=6.4$.
From the analysis of the Q$\bar {\rm Q}$ potential $V(r)$, the Abelian part $V_{\rm Abel}(r)$, and the off-diagonal part $V_{\rm off}(r)$, we find a simple but nontrivial summation formula of 
\begin{equation}
V(r) \simeq  V_{\rm Abel}(r) + V_{\rm off}(r),
\label{eq:Nontrivial}
\end{equation}
as shown in Fig.~\ref{fig:pot_Dec}(c). 
This summation formula indicates that the short-distance difference between $V(r)$ and $V_{\rm Abel}(r)$ is almost complemented by the off-diagonal part $V_{\rm off}(r)$.
Note, however, that in the non-Abelian theory this simple summation formula is fairly nontrivial, because the link variables are not commutable. 
In general, one finds 
\begin{equation*}
 {\rm tr}  \left( \prod_\mathcal{C}  M_\mu (s) u_\mu (s)  \right) 
 \not =  {\rm tr}  \left( \prod_\mathcal{C}  M_\mu (s)  \right)  
               {\rm tr}  \left( \prod_\mathcal{C}  u_\mu (s)  \right),
\end{equation*}
i.e., $W_\mathcal{C} \left[ U_\mu(s)  \right]  \not = W_\mathcal{C} \left[ M_\mu(s)  \right] \cdot  W_\mathcal{C} \left[ u_\mu(s)  \right]$, and therefore the summation formula~(\ref{eq:Nontrivial}) is nontrivial.

\section{Summary and concluding remarks}
\label{sec:Summary}
We have studied the MA projection of quark confinement in SU(3) lattice QCD by investigating the Q$\bar{\rm Q}$ potential $V(r)$, the Abelian part $V_{\rm Abel}(r)$, and the off-diagonal part $V_{\rm off}(r)$. 
We have investigated various quenched lattices 
with several spacings and volumes, e.g., $32^4$ lattice at $\beta =6.4, 6.0$, 
and have performed a potential analysis 
using both on-axis data and several types of off-axis data, 
with larger numbers of gauge configurations.
Remarkably, we have found almost perfect Abelian dominance of the confinement force (or the string tension) on the large physical-volume lattice of $32^4$ 
at $\beta =6.4, 6.0$.
In addition, we have found a nontrivial simple summation formula of 
$V(r) \simeq  V_{\rm Abel}(r) + V_{\rm off}(r)$.
Thus, in spite of the non-Abelian nature of QCD, quark confinement is entirely kept in the Abelian sector of QCD in the MA gauge.
In other words, the Abelianization of QCD can be realized without the loss of the quark-confining force via the MA projection.
This fact could be meaningful to understanding the quark-confinement mechanism in the non-Abelian gauge theory of QCD.

\begin{acknowledgments}
We thank Hideaki~Iida.
N.S. is supported by a Grant-in-Aid for JSPS Fellows (Grant No.~250588).
H.S. is supported by the Grant for Scientific Research [(C) No.23540306] 
from the Ministry of Education, Science and Technology of Japan.
The lattice QCD calculations were partially performed on NEC-SX8R at Osaka University.
This work was partially supported by RIKEN iTHES Project.
\end{acknowledgments}

\end{document}